# Maximise Global Gain in the Minority Game

Sy-Sang Liaw

*Department of Physics, National Chung-Hsing University, 250 Guo-Kuang Road, Taichung, Taiwan*

## Abstract

We find a simple, partially altruistic mechanism that can increase global gain for a community of selfish agents. The mechanism is implied in the phenomena found in the minority game. We apply the mechanism to a two-road traffic system to maximise traffic flow.

One of the central problems in many social and biological systems is how the global waste of resources by a community of agents can be reduced to a minimum. Recently, the minority game [1], inspired by the *El Farol* bar problem [2], was introduced to capture certain essential and general features of competition between adaptive agents. It has been shown that the global average gain can be larger than that of a random process only if there is enough information available to all agents [3]. Nevertheless, in real systems such as financial markets, it is more often the case that information is limited in comparison with the number of agents involved. Here we show that, even when the available information is limited, a global average gain larger than that in a random process can be achieved as long as a small but sufficient portion of agents promises to make a sacrifice – to act without considering its own benefit. The application of this finding to increasing the traffic flow in a two-road system is then discussed.

The minority game consists of *N* agents going to either room A or B based on predicting a strategy chosen from their *S* strategies. The room with the smaller number of agents is the winning room, and every agent in the winning room is awarded one point. All strategies with the correct prediction (of which room will win) also score one point. Each agent chooses his or her highest-score strategy each time. A strategy is actually a list of many entries, each prescribing which room to go to according to the information gathered from previous experiences. The standard minority game assumes that the information available to all agents is the winning history of the previous *M* (*memory*) time steps; thus, there are $2^M$ entries in each strategy. The *history* is then updated according to the current outcome of every time step.

Consider that *N* is large. If no one uses a strategy, that is, everyone randomly chooses a room, then the time sequence of the number of people in room A will form a binomial distribution centred at 0.5*N* with the standard deviation $\sigma = \sqrt{N/4}$. The average of the difference between 0.5*N* and the number of people in room A is about $0.8\sigma = 0.8\sqrt{N/4}$ [4], and the average score by this amount would be less than 0.5*N*. Thus, the smaller $\sigma$ is, the larger the global average gain. We calculate the average gain for a random process of the minority game to be $0.5 - 0.4/\sqrt{N}$. This average gain approaches 0.5 when *N* is large.

It is beyond question that people use strategies in games. Once a person has used a "good" strategy and fared better than the rest, it is certain that others will begin to develop their own "good" strategies. In a situation with limited information, however, it will soon be discovered that the average gain is less than when no one uses a strategy. The situation is much like that in the Prisoner's Dilemma [5,6], in which each agent makes the best choice for him or herself but ends up with bad results for both agents. In Fig. 1, we plot the time sequence of the number of people in room A for the process in which everyone uses his or her best strategy, which is updated at each time step, to make a decision. Apparently, the small value of the average gain in this case is due to large splits at the beginning and end of every $2^{M+1}$ interval of the time steps. We explained recently [7] that this quasi-periodic structure is due to the form of the payoff function and the way in which everyone uses his or her strategy. In the case of a large number of agents, one way to raise the long-term global average gain is to introduce a learning mechanism [1], by which bad performers partially replace their strategies with better ones, or certain genetic adaptation schemes whereby each agent modifies parts of his or her bad strategies based on better ones [8]. Here we show two ways of achieving a high global average gain in the standard minority game – all agents continue to use their best strategies from among the *S* strategies they selected at the beginning of the game without modifying or substituting them.

First, assume that every agent is reluctant to change strategy unless there is another at hand that has proved to be much better – having gained $l \gg 1$ points more than the current one. In this *conservative* community, the reluctance to switch to another strategy offers strong stability against variation. As a result, the number of people in room A gradually grows closer to *N*/2 so that the global gain increases. One might think the reason for this is that a strategy with predictions distributed more evenly between rooms A and B will score better and thus be favoured. However, this is not true. We have found that when the number of agents converges to *N*/2 as time goes on, the variance of the number of predictions in room A for all of the best strategies remains more or less constant. What really occurs is that, because the winning sequence is generally set according to the persistent use of strategies, there are only a few agents who change strategies at each time step, which means that the variance of the predictions hardly changes. At the same time, the act of switching to a new strategy increases by a small amount the number of agents in the winning room when the same pattern occurs again, thus bringing the number of agents in that room closer and closer to *N*/2. Consequently, the global average gain increases as time goes on (see the curve indicated by *conservative* in Fig. 2). As we have pointed out in our previous work [9], the crucial factor in reducing population variance is to decrease the number of agents who switch strategies at the same time. The simplest way to accomplish this is to allow only a certain number of agents to switch strategies at a given time [9]. In a real system, such as the stock market, such a limitation measure would be totally unacceptable to agents. However, we have found that a society of conservative agents automatically reduces the number of simultaneous switches. The method works when agents are conservative only at the beginning – before their first strategy switch. In a simulation, we allow one of an agent's strategies to have $l$ points and the others to have 0 points at the beginning [10,11], and the game is played as usual – all agents use their highest-score strategy each time. To understand why this method works, let us consider the simplest case of $S = 2$. There are at least $l$ time steps before the first strategy switch is made. During this period, because everyone's favourite strategy is chosen randomly at the beginning, the score differences between the two strategies disperse among agents, and their distribution is given approximately by a normal distribution centred at $l$. The probability of the score difference being zero is small because it is at one tail of a normal distribution. This can be compared with the case of a standard game ($l = 0$) in which the probability of the score difference being zero is at the maximum of a normal distribution. It is clear that, according to the rule of using the highest-score strategy, a switch can be made at one particular time step only when the score difference between the two strategies of a given agent was zero at the previous time step. Consequently, switching probability is greatly reduced when $l$ is large.

There is a second method of increasing global gain that works without requiring agents to be conservative. Suppose the prediction of each entry in every strategy has a $p \leq 0.5$ probability of being 0 (room A) and (1-$p$) of being 1 (room B) [12,13]. A random prediction means $p = 0.5$. Now, choose $p$ to be slightly smaller than 0.5, for example, 0.4825, in a case with $M = 8$ and $S = 2$. A typical time sequence for people in room A is shown in Fig. 3. We see that after a short period of time, the variation of this sequence begins to decrease so that the global average gain increases (see the curve indicated by *partially altruistic* in Fig. 2). Now, let us consider *p* to be the percentage of 0s in all entries of the strategies. Suppose that 3.5% of agents always choose room B, and the rest use their best strategies as usual [14]. This is equivalent to setting $p = 0.4825$ ((1-0.035)/2 = 0.4825). In this way, the mechanism does not require that all agents behave in the same particular way, for example, by acting conservatively. What we need is for a small portion of agents to act unselfishly (to forgo their best strategies so that their behaviour is considered to be *altruistic*); then, the global gain improves in a self-organised manner. Compared to the first method, this mechanism is more likely to operate in real situations, either because of the altruistic behaviour of some agents (which has been observed in many biological species) or because of the careful arrangements of the game coordinator (such as government).

Two stages are required for the second mechanism to produce the result shown in Fig. 3. First, the average number of people in room A has to shift upward from *pN* to become closer to *N*/2. Second, the time sequence reduces its standard deviation as time goes on. The reason the average value shifts upward is as follows [7]. In the first $2^M$ time steps, room A always wins. Thus, at the end of $2^M$ time steps, the strategies that have the most 0s score best and will be used in the following time steps. The average number of 0s for each agent's best strategies is larger than the average of all strategies, which is $2^M p$. Therefore, the average number of people in room A shifts upward [7]. If this shift is well-adjusted [15] – for the subsequent time steps, room A has about a 50% chance of winning – then, because these best strategies have already accumulated more points than the others, the situation at time step $2^M + 1$ is similar to that in the first method in which each agent is initially biased in favour of one strategy. Consequently, the standard deviation of the time sequence begins to decrease.

It is worth mentioning that the mechanisms work even better when *S* is larger. When *S* is larger, the average gain will be smaller if everyone uses his or her best strategy in the standard game. However, the average gain can still be brought very close to 0.5 if a small but sufficient number of people [16] promise to make a sacrifice. It is also important that the mechanism works for the whole range of values of $\rho = 2^M / N$ (see Fig. 4). The effect of the gain-increasing mechanism is most efficient in the small $\rho$ case in which global gain is low in the standard game.

The *partially altruistic* mechanism described above can be employed in a two-road traffic system. There are two express highways running from south to north in Taiwan. People who work in Taipei, the capital of Taiwan, normally drive south to their parents' homes or recreation destinations on the first day of a long holiday and drive back north on the last day. Traffic jams are a nightmare for everyone, and thus the good choice of a highway from the two available is crucial to the holiday mood.

Highway traffic flow has been extensively studied [17]. Recently, the cellular automata approach [18] has obtained a number of interesting results. The fundamental diagram in traffic models is the plot of traffic flow as a function of vehicle density. A typical fundamental diagram shows that traffic flow reaches a maximum at a certain vehicle density and that the curve as a function of density is concave-upward. Thus, for a system with two similar roads, as in the case of Taiwan, the total traffic flow would be maximal when vehicles are distributed evenly on both. The problem of maximising traffic flow in a two-road system is therefore similar to maximising the global average gain in the minority game, which aims to distribute agents evenly in two rooms. Assume that every driver has a few strategies for determining which road to take, and that on every occasion he or she chooses the strategy that most often correctly predicted the road with the better flow in the past [19]. According to the second mechanism discussed above, one possible way of obtaining a better traffic flow for a given number of vehicles is to have a few percent [16] drivers choose the same highway every time, either voluntarily or by design.

Are those who do not adopt a strategy really sacrificing anything? In our simulation, they gain no less in the long run. What they have really sacrificed is the freedom to choose.

**Acknowledgments**

This work was supported by grants from the National Science Council (grant number NSC93-2112-M005-002) and the National Center for Theoretical Sciences in Taiwan.

**Figure captions**

Fig. 1

Time sequence of the population in room A in the standard minority game for $N = 10001$, $M = 8$, and $S = 2$. The central line shows the position of 0.5$N$. A quasi-periodic structure can be readily seen. At the ends of each quasi-periodic interval, the population has a large deviation from the central line.

Fig. 2

The average gain of each $2^{M+1}$ time step as a function of time. *Naïve* (dots): no agent uses a strategy; *Selfish* (diamonds): every agent uses his of her best strategy; *Conservative* (crosses): agents are reluctant to change strategies at the beginning by giving $l = 30$ points more to one strategy; *partially altruistic* (circles): 3.5% of agents ignore their best strategies and make constant choices ($N = 10001$, $M = 8$, $S = 2$).

Fig. 3

When 3.5% of agents always choose room B and the others use their best strategies, the number of people in room A converges to $N/2$ as time goes on. The lower short line shows the position at $0.4825N$ ($N = 10001$, $M = 8$, $S = 2$).

Fig. 4

Global average gain as a function of $\rho = 2^M / N$ ($M = 8$, $S = 2$). The results for the *partially altruistic* case (solid curve) are always better than those for the *naïve* (random) case (dotted) or the *selfish* case (circled). The latter shows a maximum at some $\rho$ and is better than the *naïve* case only when $\rho$ is large enough. (The results are the average of 30 samples of random initial strategies. For each sample, we ran $50 \cdot 2^M$ time steps and collected the data of the last $2^{M+1}$ time steps for the average.)

Fig. 1

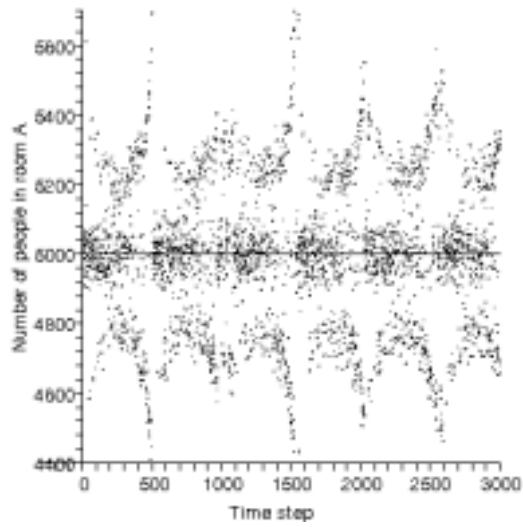

Fig. 2

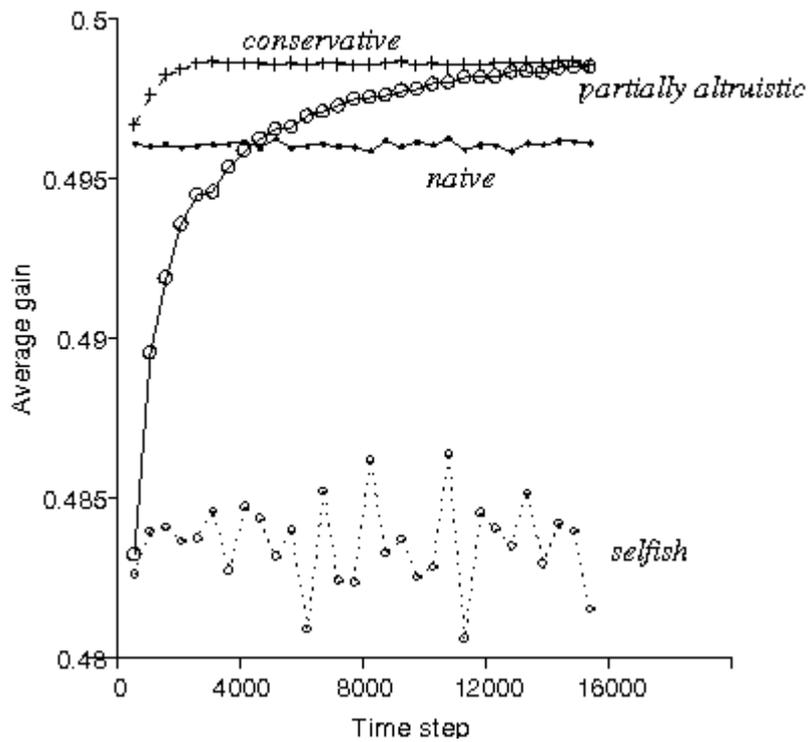

Fig. 3

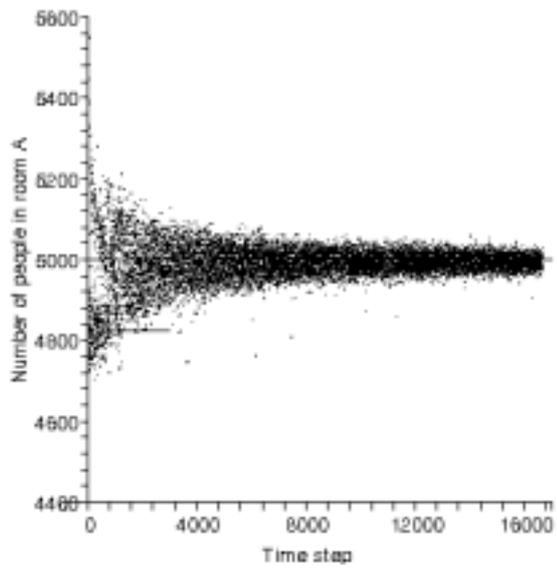

Fig. 4

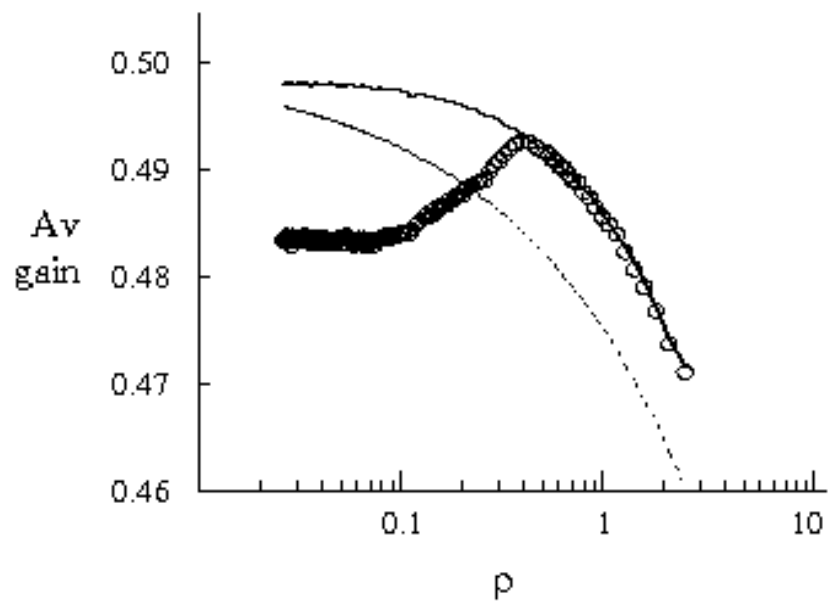